\documentclass[
preprint, 
 amsmath,amssymb,
 aps, 
pra,
]{revtex4-2}

\usepackage{graphicx}
\usepackage{dcolumn}
\usepackage{bm}
\usepackage[dvipsnames]{xcolor}

\begin{document}

\preprint{APS/123-QED}

\title{\textbf{A contaminant-concentration-dependent surface tension does not explain the absence of solutal Marangoni flow in evaporating droplets} 
}%

\author{Javier Martínez-Puig}
\email{jmpuig@pa.uc3m.es}

\author{Théophile Gaichies}
\author{Javier Rodríguez-Rodríguez}

\affiliation{Carlos III University of Madrid, Department of Thermal and Fluids Engineering, Leganés, Spain}

\date{\today}

\begin{abstract}
Theoretical models of evaporating droplets predict Marangoni flows orders of magnitude faster than those observed experimentally. While this discrepancy is often attributed to surface contamination, the underlying mechanism by which contaminants weaken Marangoni stresses remains unclear. In this study, we compare particle image velocimetry (PIV) experiments with a coupled hydrodynamic and solute transport model to investigate the internal flow of evaporating aqueous droplets containing salt, glycerol, or ethanol. By analyzing both sessile and pendant droplets, we demonstrate that the flow is driven entirely by natural convection, in contrast to theoretical predictions that use surface-tension gradients. Remarkably, in some cases, the experimental surface velocity is found to be directed against the predicted surface-tension gradient. We further prove that commonly used contamination models—whether based on surfactants lowering the surface tension or on surface rheology—cannot account for this flow reversal, as they would require the surfactant to be transported against the flow when the surface velocity is opposed to the surface-tension gradient. Our results show that a reconsideration of these models is required to explain the absence of solutal Marangoni in evaporating droplets.
\end{abstract}

\maketitle

Since the seminal works of the Chicago group \cite{Dee} and those of Hu and Larson \cite{HuLarson}, it has been established that in an evaporating droplet, the measured magnitude of Marangoni convection—whether thermal or solutal—frequently disagrees by orders of magnitude with theoretical predictions. Using surface-tension variations due to temperature or solute concentration reported in the literature, the predicted Marangoni velocity is typically on the order of $\mathrm{mm/s}$, whereas experimental measurements are on the order of $\mathrm{\mu m/s}$. Although good quantitative agreement has been reported in some studies \cite{Kazemi2021_ThMar}, this discrepancy remains widely documented in millimetric water droplets evaporating under ambient conditions for thermal Marangoni flow \cite{HuLarson, XuLuoMarWat, TrantumMarWat} and for multicomponent aqueous solutions in the case of solutal Marangoni flow \cite{MarinStain, martinezpuig2025}, and has been attributed to the presence of surface contaminants that counteract surface tension gradients \cite{HuLarson, KangRayleighSaline, GelDiddMar, Rocha_SolutMarRayleigh}.

In a multicomponent aqueous droplet, evaporation generates concentration gradients. Consequently, three mechanisms can, in principle, be present in the flow: capillary flow, Marangoni convection arising from surface-tension gradients, or natural convection driven by density gradients. Thermal effects, induced by evaporative cooling, can be ruled out at ambient conditions provided solutal gradients dominate surface tension and density variations. For droplets with contact angles smaller than $90^\circ$, the capillary flow is directed from the apex of the droplet toward the contact line. This implies that whenever recirculation is present in the droplet, it must be due to either Marangoni or natural convection. Although Edwards et al. \cite{EdwardsDensityDriven} and Li et al. \cite{LiGravEffe} showed that gravity can be a driving force in evaporating droplets even when the Bond number is small—though neither study provides a conclusive explanation for why Marangoni convection is absent—many studies still identify recirculation exclusively with Marangoni convection—notably in thin droplets, even when measured velocity fields are on the order of magnitude predicted by natural convection \cite{MarinStain, EfstratiouCryDrivFlows, BasuFomitesBact}. This oversight typically originates from the assumption that, in the limit of small droplet height, the vertical diffusion timescale is sufficiently short to ensure rapid solutal homogenization in that direction. However, this does not imply that gravitational forces can be neglected. Indeed, as demonstrated by Selva et al. \cite{SelvaSalmonNatConvPRL}, natural convection can effectively drive a recirculatory flow in an evaporating confined droplet even when the concentration gradients are oriented orthogonal to the gravitational force. Consequently, the presence of recirculation in thin, multicomponent droplets cannot be attributed a priori to Marangoni stresses alone.

The discrepancy between theoretical and experimental velocities is frequently attributed to surface contamination, a phenomenon particularly prevalent in polar liquids such as water. Despite the well-documented presence of such contaminants at the air-water interface \cite{MizevThresholdMarangoni, PonceTorres_OsLiBri, MoaleiPRL_IncInt, Franiatte_PRL2021}, the precise mechanism by which they suppress Marangoni convection remains elusive. In this Letter, we demonstrate that widely used contamination models—specifically those involving soluble or insoluble surfactants that monotonically reduce surface tension, or rheological models based on the Boussinesq-Scriven law— predict surface velocities incompatible with our experimental findings. In particular, when the measured surface velocity is opposed to the surface-tension gradient, we show that, to account for this flow reversal, the surfactant would have to be transported against the flow, leading to a physical contradiction.

\begin{figure*}[ht!]
    \centering
    \includegraphics[width=\textwidth]{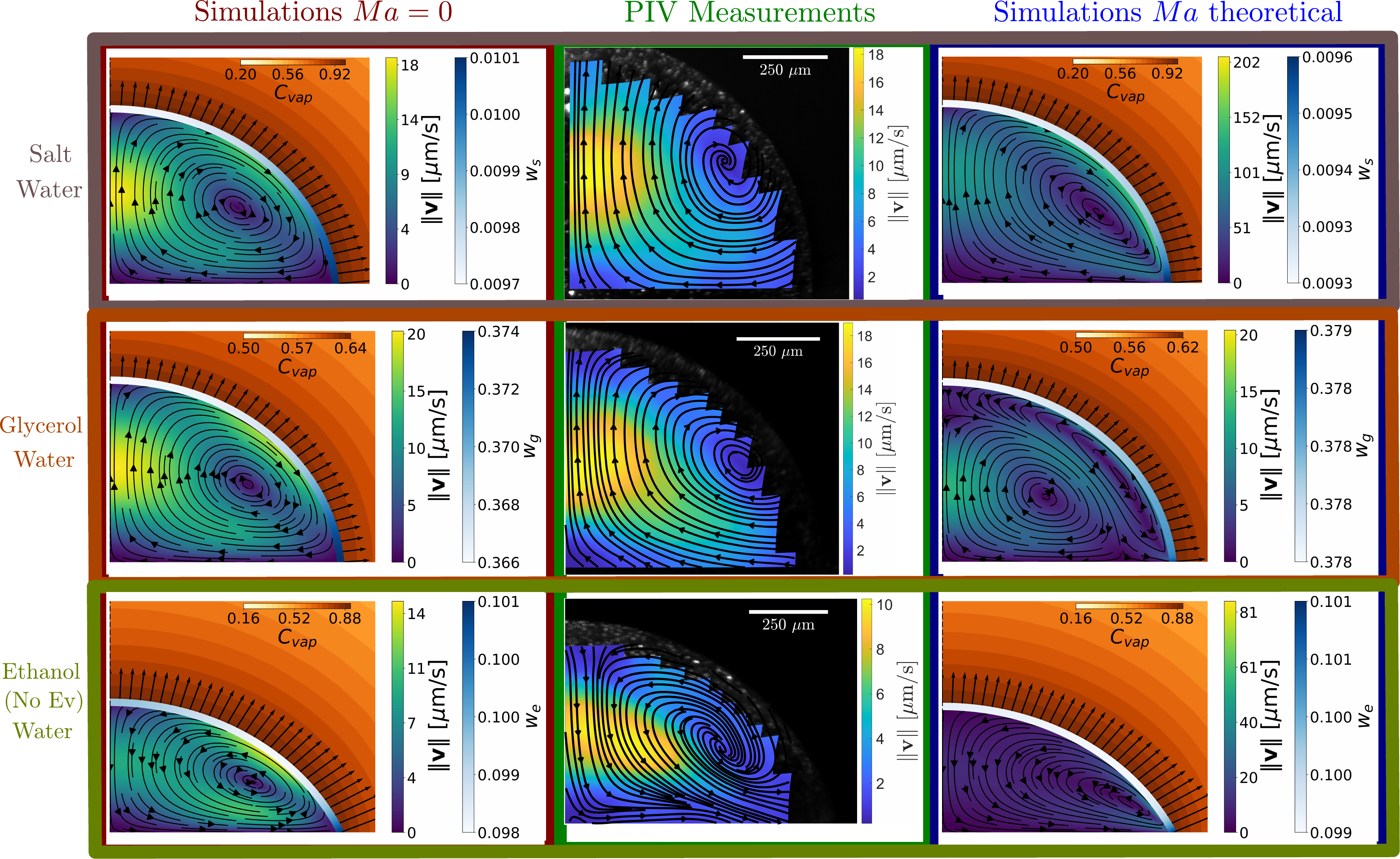}
    \caption{Evaporation of sessile salt--water, glycerol--water, and ethanol--water droplets.  The initial volumes, contact radii and relative humidity are, from the top to the bottom row:  $V_0 = 0.72 \ \mu\text{L}$, $R_c = 0.73 \ \text{mm}$, $H_r=20\%$;  $V_0 = 0.94 \ \mu\text{L}$, $R_c = 0.79 \ \text{mm}$, $H_r=50\%$;  $V_0 = 0.89 \ \mu\text{L}$, $R_c = 0.86 \ \text{mm}$, $H_r=20\%$. In the case of ethanol--water droplets the gas phase is saturated with ethanol vapor to avoid ethanol evaporation. Simulations performed without surface–tension gradients ($\mathrm{Ma}=0$) show good agreement with experimental measurements, whereas simulations that include surface–tension variations do not.}
    \label{fig:Ma0PIVMaThe}
\end{figure*}

\emph{Methods.} To investigate the discrepancy between experimental and theoretical velocity fields in multicomponent droplets, we developed a detailed mathematical model of the evaporation-induced flow within the droplet. While a comprehensive description of the model is provided in \cite{martinezpuig2025}, we highlight its primary features here. The transport of water vapor in the surrounding air is modeled as a purely diffusive process. Within the droplet, inertia is neglected, and we solve the Stokes equations with buoyancy forces, which read in dimensionless variables as
\begin{equation}
\boldsymbol{\nabla}\cdot\tau - Ra\,w_s\,\boldsymbol{e_z} = 0,
\end{equation}
where $\tau = -P\mathcal{I} + (\boldsymbol{\nabla}\boldsymbol{v} + \boldsymbol{\nabla}\boldsymbol{v}^T)$ is the dimensionless stress tensor defined using the reduced pressure $P$, and $Ra = R_c^3 g \rho_s / \mu D_s$ is the solutal Rayleigh number. Here, $R_c$ is the droplet contact radius, $g$ is the gravitational acceleration, $\rho_s = \partial\rho/\partial w_s$, $\mu$ is the liquid dynamic viscosity, and $D_s$ is the solute diffusivity. 

Marangoni stresses are incorporated via the dimensionless tangential stress condition at the air-water interface:
\begin{equation}
\boldsymbol{n}\cdot\tau\cdot\boldsymbol{t} = Ma_s\boldsymbol{\nabla_{t}}w_s\cdot\boldsymbol{t},
\end{equation}
where $\boldsymbol{n}$ and $\boldsymbol{t}$ are the unit normal and tangential vectors to the interface, $\boldsymbol{\nabla_{t}}$ is the surface gradient operator, and $Ma_s = \gamma_s R_c / \mu D_s$ is the solutal Marangoni number, with $\gamma_s = \partial\gamma/\partial w_s$. Density and surface tension variations are assumed to vary linearly with the solute mass fraction $w_s$.

The fluid flow is coupled to advective-diffusive solute transport within the droplet, governed by
\begin{equation}
\partial_t w_s + \boldsymbol{v}\cdot \boldsymbol{\nabla}w_s = \nabla^2 w_s.
\end{equation}
All equations are transformed into their weak form and solved using COMSOL Multiphysics.

We evaporated aqueous droplets ($V_0 = 0.8 \pm 0.1\,\mu\mathrm{L}$) containing sodium chloride, glycerol, or ethanol in sessile and pendant modes to span a range of solutal effects on surface tension and density. To ensure pure water evaporation in ethanol mixtures, the atmosphere was saturated with ethanol vapor. Internal velocity fields were measured via PIV using a thin laser sheet and $1.11\,\mu\mathrm{m}$ polystyrene tracers. Light refraction at the interface was corrected using ray tracing \cite{KwanRayTracing}, while interfacial velocities were obtained by averaging over quasi-stationary periods. This technique is enabled by the fact that the velocity at which the interface recedes is very small compared with the velocities inside the drop. Notably, the primary value of these interfacial measurements lies in establishing the flow direction. This qualitative feature remains reliable despite experimental noise, as corroborated by the observation of particles straddling the interface whose displacements align with PIV measurements.

\emph{Missing Marangoni convection.} In our experimental measurements, we consistently observed a recirculatory flow for all three solutions evaporated in both sessile and pendant modes. To elucidate whether the primary mechanism driving this recirculation is Marangoni or natural convection, we solved our model using material properties taken from the literature \cite{NaClSurfTens, EthanolSurfTen, GlycerolPhysProp}. 

In Fig.~\ref{fig:Ma0PIVMaThe}, we show snapshots of experiments alongside simulations conducted both with ($\mathrm{Ma}$ theoretical in Fig.~\ref{fig:Ma0PIVMaThe}) and without ($\mathrm{Ma}=0$ in Fig.~\ref{fig:Ma0PIVMaThe}) Marangoni stresses. Despite differences in solute properties and evaporation modes, we consistently found that the flow is quantitatively reproduced only when Marangoni effects are neglected, as was the case for Kang et al. \cite{KangRayleighSaline} for saline droplets. This holds even though Marangoni effects should be dominant when the actual values for surface-tension variations with solute concentration, reported in the literature, are used in the simulation (see Fig.~\ref{fig:Ma0PIVMaThe}).

\begin{figure*}[ht!]
    \centering
    \includegraphics[width=\textwidth]{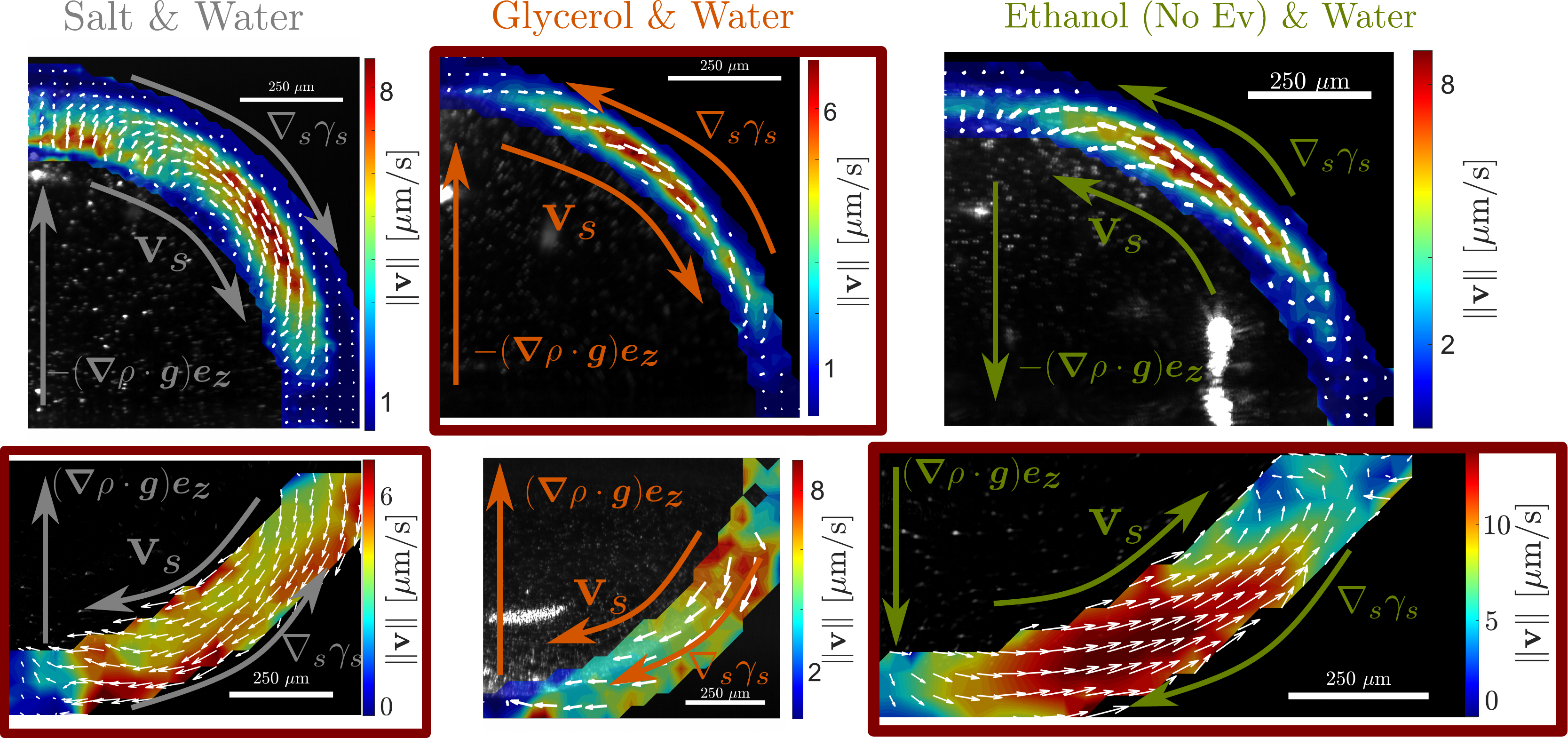}
   \caption{Averages of 10 images ($\sim 4$ s) of interfacial velocities for evaporating salt--water, glycerol--water, and ethanol--water droplets in sessile and pendant modes. Red rectangles highlight experiments where density and surface tension gradients point in opposite directions. In these experiments, the surface velocity follows the density gradient and opposes the surface tension gradient.}
    \label{fig:IntVelExpAll}
\end{figure*}

Interestingly, including surface-tension variations in the simulations increases the discrepancy with experimental data by at least one order of magnitude and, in several cases, fails to predict the correct flow direction. For instance, in a sessile glycerol--water droplet, evaporation is stronger near the contact line. This leads to glycerol accumulation in that region, creating a surface-tension gradient directed toward the apex. Such a gradient would reverse both the experimentally observed flow and the flow predicted by natural convection alone (see Fig.~\ref{fig:IntVelExpAll}).

Furthermore, to verify that the driving force behind the convection is indeed gravitational, we evaporated the same solutions in the pendant drop configuration to inverse the direction of gravity in the referential of the drop (see Fig.~\ref{fig:IntVelExpAll}). For all three solutions, we observed that the direction of the velocity field is reversed. The droplet's interface remains a spherical cap due to the small Bond number. Thus, in the pendant drop configuration the only difference is the direction of the buoyancy forces; therefore, the observed reversal of the flow confirms that the recirculation is driven by buoyancy effects. We note that the reversal of the recirculation direction between the pendant and sessile configurations rules out the possibility that thermal Marangoni effects associated with evaporative cooling are driving the flow.

To understand why modeling the effect of contaminants simply as a reduction in the surface tension
fails to explain the absence of Marangoni convection, it is essential to study the experimental interfacial velocity. We focus specifically on cases where the density and surface-tension gradients point in opposite directions: sessile glycerol--water droplet, and pendant salt--water and ethanol--water droplets (see red panels in Fig.~\ref{fig:IntVelExpAll}). In these three cases, the experimental surface velocity is aligned with the density gradient and opposite to the surface-tension gradient.

\emph{Current models of surface contaminants predict surface velocities incompatible with our experimental findings} The mismatch between theoretical and measured velocities has often been attributed to surface contamination from unknown sources. Such contamination can significantly alter the flow even at concentrations too low to be detected by conventional surface-tension measurement techniques \cite{HuLarson, KangRayleighSaline, GelDiddMar, Rocha_SolutMarRayleigh}. Rather than focusing on the specific origin of such contamination, we investigate its potential implications for the flow field. Surface contamination can influence the droplet through surface tension changes or via energy dissipation at the surface; these are typically modeled using insoluble surfactants and the Boussinesq–Scriven law \cite{scriven1960dynamics,boussinesq1913speed}, respectively. While Rocha et al. \cite{Rocha_SolutMarRayleigh} focused on the former, the rheological effects of such contamination remain largely understudied in evaporating droplets.

Consider an insoluble surface contaminant at a low concentration $\Gamma$ that linearly reduces the local surface tension. Its transport is governed by the surface advection--diffusion equation; the surface tension is modified to first order as: $\gamma(w_s, \Gamma) = \gamma_s(w_s) + \gamma_\Gamma \Gamma$ where $\gamma_s$ is the surface tension of the mixture as a function of solute mass fraction $w_s$ without any surfactant, and $\gamma_\Gamma$ is the linear coefficient describing the reduction of surface tension due to the surfactant. For the sake of the argument, we further assume that the variation of surface tension in the absence of surfactant is linear in the solute mass fraction, $\gamma_s(w_s) = \gamma_0 + \gamma_s w_s$, where $\gamma_0$ is the surface tension of pure water. However, our reasoning can extend to any, more realistic, constitutive equation $\gamma(w_s, \Gamma)$ as long as it is monotonic in $\Gamma$.

\begin{figure*}[ht!]
    \centering
    \includegraphics[width=\textwidth]{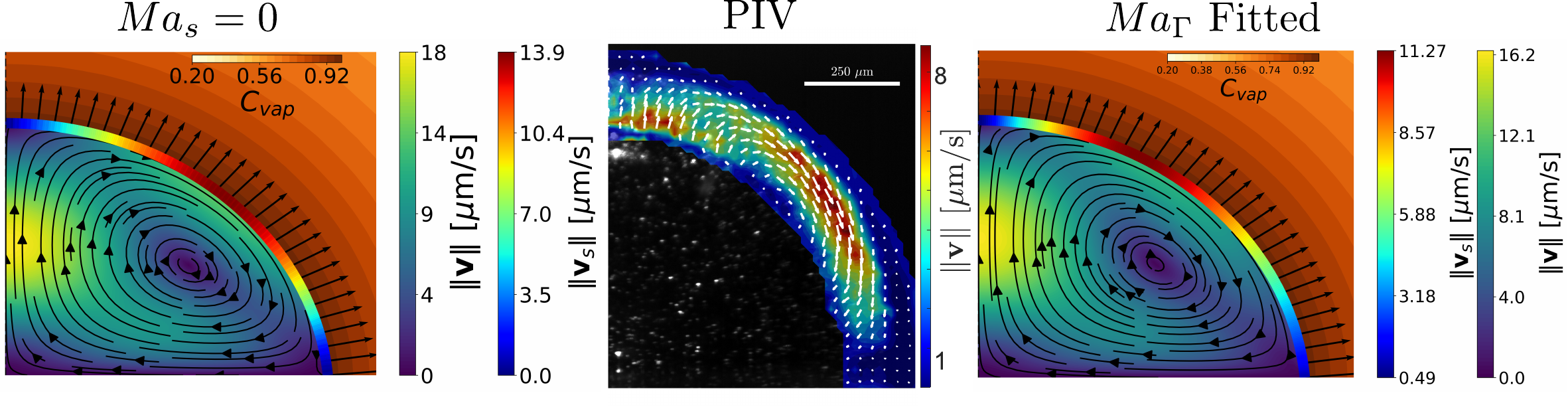}
    \caption{Evaporation of a sessile salt--water droplet. Simulations without surfactants or surface-tension gradients show good agreement with experiments ($\mathrm{Ma}_s=0$). In this case as the surface-tension gradient and the surface velocity are parallel, surfactants can be included to counteract the salt-induced gradient. By fitting the surfactant-associated Marangoni number, we successfully reproduce the experimental observations ($\mathrm{Ma}_\Gamma$ Fitted).}
    \label{fig:IntVelSalt}
\end{figure*}

The balance of tangential stresses at the interface reads 

\begin{equation}\label{eq:TanStress}
\mathbf{n} \cdot \mathbf{\tau} \cdot \mathbf{t} = \nabla_s \left(Ma_s w_s + Ma_\Gamma \Gamma \right) \cdot \mathbf{t},
\end{equation}

and is governed by Marangoni numbers, $\mathrm{Ma}_s= \gamma_s R_c / (\mu D_s)$ and $\mathrm{Ma}_\Gamma = \gamma_\Gamma R_c / (\mu D_s)$, which quantify the surface-tension sensitivity to solute and surfactant concentrations, respectively. The sign of the Marangoni number depends on whether the solute or surfactant concentration increases (positive) or decreases (negative) the overall surface tension. Van Gaalen et al.~\cite{DiddnesSurf} demonstrated that in the limit $|\mathrm{Ma}_\Gamma| \to \infty$, the interfacial velocity $\mathbf{v_s} \to 0$. Rocha et al.~\cite{Rocha_SolutMarRayleigh} further showed that even small reductions in surface tension due to surfactants can substantially reduce interfacial velocities, effectively suppressing solutal or thermal Marangoni effects. While our simulations confirm this behavior, our experiments reveal non-negligible interfacial velocities (see Fig.~\ref{fig:IntVelSalt}). Theoretically, it remains possible to reproduce the experimental velocities by fitting $\mathrm{Ma}_\Gamma$, as long as the surface-tension gradient and the surface velocity are in the same direction. For instance, by minimizing the difference between the experimental and numerical velocities via a least-squares approach, we find $\mathrm{Ma}_\Gamma = -1.63 \times 10^5$ for a sessile NaCl--water droplet. Although this fitting procedure is somewhat arbitrary ---as it requires a precise surfactant-induced reduction in surface tension--- it allows us to quantitatively reproduce both bulk and interfacial velocity fields (see Fig.~\ref{fig:IntVelSalt}).

\begin{figure*}[ht!]
    \centering
    \includegraphics[width=\textwidth]{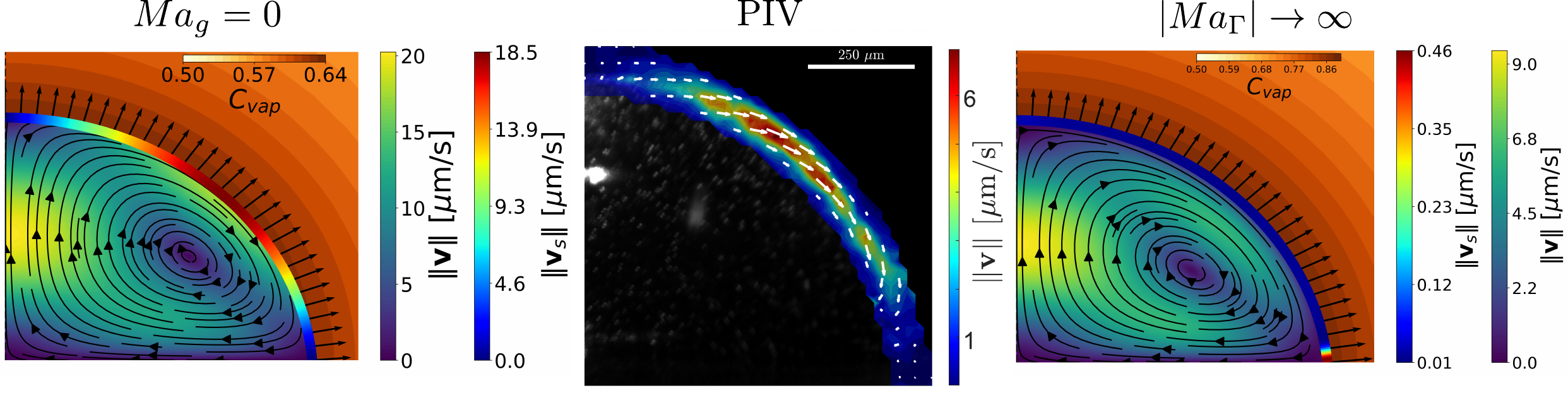}
   \caption{Evaporation of a sessile glycerol--water droplet. Simulations without surfactants or surface-tension gradients show good agreement with experiments ($\mathrm{Ma}_g=0$). In this case as the surface velocity and the surface-tension gradient are anti-parallel, surfactants fail to reproduce the experimental observations. Even at extremely high surfactant Marangoni numbers ($|\mathrm{Ma}_\Gamma|\to\infty$), the positive surface radial velocities observed experimentally cannot be achieved; instead, the interfacial velocity simply approaches zero.}
    \label{fig:IntVelGlyc}
\end{figure*}

However, this picture changes significantly when the surface-tension gradient and the surface velocity point in opposite directions, as is the case for sessile glycerol--water droplets. This behavior is fundamentally incompatible with the classical idea that Marangoni flows are suppressed by surfactant-induced surface-tension reduction. We have verified numerically that even in the limit $|\mathrm{Ma}_{\Gamma}|\to\infty$, where the interfacial velocity vanishes ($||\mathbf{v_s}||\to0$), the flow is never reversed (see Fig.~\ref{fig:IntVelGlyc}). 

For the sake of the argument, we consider a thin droplet within the lubrication approximation (where the droplet height, $h$, is much smaller than its contact radius, $R_c$). However, the reasoning remains the same even if the drop was not thin enough to apply lubrication. In this framework, the tangential stress condition at the interface reads

\begin{equation}
    \frac{\partial v_r}{\partial z} = \mathrm{Ma}_g \frac{\partial w_g}{\partial r} + \mathrm{Ma}_\Gamma \frac{\partial \Gamma}{\partial r}.
\end{equation}

In a glycerol--water droplet, both $\mathrm{Ma}_g$ and $\mathrm{Ma}_\Gamma$ are negative. One might initially assume that if the surfactant concentration exactly cancels the solutal Marangoni stress induced by the glycerol gradient, it would yield a stress-free condition at the interface, analogous to setting $\mathrm{Ma}_\Gamma = \mathrm{Ma}_s = 0$. However, this assumption imposes specific requirements on the surfactant distribution that lead to a physical contradiction. To counteract the solutal Marangoni stress induced by the glycerol concentration, the surfactant gradient must satisfy:
\begin{equation}
    \frac{\partial \Gamma}{\partial r} = -\frac{\mathrm{Ma}_g}{\mathrm{Ma}_\Gamma} \frac{\partial w_g}{\partial r} < 0,
\end{equation}
since the non-uniform evaporation rate produces $\partial w_g / \partial r > 0$ and the ratio $\mathrm{Ma}_g / \mathrm{Ma}_\Gamma$ is positive. We can assume for the sake of simplicity that the surfactant transport is quasi-steady. For a thin droplet curvature effects in the surface advection--diffusion equation can be neglected, and the conservation of surfactant at the interface reduces to:
\begin{equation}
    \frac{1}{r} \frac{\partial}{\partial r} \left( \Gamma\, r\, v_r \right) = \mathrm{Pe}_\Gamma^{-1} \frac{1}{r} \frac{\partial}{\partial r} \left( r \frac{\partial \Gamma}{\partial r} \right),
\end{equation}
where $\mathrm{Pe}_\Gamma = v_c R_c / D_\Gamma=D_s/D_\Gamma$ is the surface Péclet number, representing the ratio of advective to diffusive surfactant transport. Integrating this equation and applying the axisymmetry condition ($\partial \Gamma / \partial r = 0$ at $r=0$) yields:
\begin{equation}
    \frac{\partial \Gamma}{\partial r} = \mathrm{Pe}_\Gamma\, \Gamma\, \left. v_r \right|_{z=h} > 0,
\end{equation}

provided that $\mathrm{Pe}_\Gamma, \Gamma, v_r>0$. This result directly contradicts the condition required to suppress the solutal Marangoni effect in glycerol--water droplets, which, as established previously, would require a negative surfactant gradient ($\partial \Gamma / \partial r < 0$). Physically, the contradiction arises because explaining the absence of Marangoni convection through this model would require the contaminants to be transported against the direction of the flow.

Consequently, when the surface-tension gradient and the observed surface velocity are in opposite directions, the presence of an insoluble surfactant that reduces surface tension cannot explain the observed reversal of the interfacial flow. Furthermore, reinforcing this conclusion, we find that the experimental interfacial velocities are accurately reproduced by simulations in which Marangoni stresses are entirely neglected, i.e., $\mathrm{Ma}_s = 0$ (see Figs.~\ref{fig:IntVelSalt} and \ref{fig:IntVelGlyc}).

The consideration of a soluble surfactant introduces additional complexity, but it does not alter the overall picture. For a water--glycerol droplet, a soluble surfactant could in principle change the orientation of the Marangoni convection if $\partial \Gamma / \partial r < 0$. To avoid the inconsistencies found for an insoluble surfactant, such a condition would need to arise not from interfacial advection of surfactant, but from enhanced adsorption near the apex of the droplet. However, the bulk surfactant concentration is higher near the contact line due to stronger evaporation in that region. Therefore, in the limit of very slow interfacial compression ($\mathbf{v}_{I}\cdot\mathbf{n} \sim 0$), there is no clear mechanism that would produce $\partial \Gamma / \partial r < 0$ other than interfacial advection of surfactant, which would lead to the same inconsistency than in the case of an insoluble surfactant.

Surface rheology, modeled by the Boussinesq--Scriven law, also fails to resolve the discrepancy. This model treats the interface as a 2D Newtonian fluid where surface shear ($\mathrm{Bq}$) and dilatational ($\mathrm{Bq}_\kappa$) Boussinesq numbers represent the ratio of surface to bulk viscous stresses. For a thin droplet the tangential stress balance is:
\begin{equation}
\frac{\partial v_r}{\partial z} = \mathrm{Ma}_g \frac{\partial w_g}{\partial r} + (\mathrm{Bq} + \mathrm{Bq}_\kappa) \frac{\partial}{\partial r} \left[ \frac{1}{r} \frac{\partial (r v_r)}{\partial r} \right].
\end{equation}
In the limit of large surface viscosity required to suppress Marangoni flow, the interface must satisfy a nearly shear-free condition:
\begin{equation}
\frac{\partial}{\partial r} \left[ \frac{1}{r} \frac{\partial(r v_r)}{\partial r} \right] = -\frac{\mathrm{Ma}_g}{\mathrm{Bq} + \mathrm{Bq}_\kappa} \frac{\partial w_g}{\partial r}.
\end{equation}
For a sessile glycerol droplet, where $\mathrm{Ma}_g > 0$ and $\partial w_g / \partial r > 0$, the integration of this expression with no flux boundary conditions ($v_r = 0$ at $r=0, 1$) yields a characteristic inward interfacial flow ($v_r \le 0$). This directly contradicts the outward radial velocity observed in our experiments. Consequently, surface rheology cannot account for the observed flow reversal.

Thus, as with the case of an insoluble surfactant, attempting to recover a shear-free condition by invoking surface viscosity necessitates an interfacial radial velocity that is negative---directly contradicting our experimental observations for sessile glycerol--water droplets. Analogous arguments hold for pendant salt--water and ethanol--water droplets, where the surface-tension gradient and the measured surface velocity also point in opposite direction (see red panels in Fig~\ref{fig:IntVelExpAll}). This leads to the surprising conclusion that while the suppression of Marangoni flows is conventionally attributed to surface contamination, neither the reduction of surface tension by contaminants nor the rheological effects described by the Boussinesq--Scriven law can explain the observed reversal of the interfacial velocity relative to the surface tension gradient.

\emph{Concluding remarks.} We do not intend to question the existence of solutal Marangoni convection in evaporating droplets. In fact, several studies have reported solutal Marangoni flow velocities on the order of millimeters per second. Christy et al. \cite{ChristyWatEth} and Diddens et al. \cite{DiddensWatEth} measured such velocities in water–ethanol droplets, while Raju et al. \cite{RajuMarRingWatGlyc} reported comparable values in water–glycerol droplets and Ramírez-Soto et al. \cite{RamirezSoto12PD} in water–12-propanediol droplets. Notably, none of these studies describe cleaning procedures that deviate substantially from standard laboratory practice, further complicating the picture we have presented throughout this work.

We emphasize that the question of how surface-active contaminants might suppress surface-tension gradients remains unresolved. We show that assuming contaminants produce a surface counter-gradient leads to a surface flow direction incompatible with experimental observations. Our results show that simply assuming negligible surface-tension gradients in the model—while retaining buoyancy effects—is sufficient to reproduce all experimental observations. This raises a fundamental question: why, then, is Marangoni convection so rarely observed in evaporating droplets? Any attempt to answer this question must go beyond the conventional models typically used to characterize surface contamination.

\emph{Acknowledgments.} We would like to thank Christian Diddens, Álvaro Marín and Duarte Rocha for their insightful discussions on a preliminary draft of this work. We would like to especially acknowledge Christian Diddens for suggesting the experimental measurement of these solutions in pendant drops. The authors acknowledge financial support from Grant No. PID2023-146809OB-I00 funded by MICIU/AEI/10.13039/501100011033.

\bibliography{apssamp}

\end{document}